# and Solitonic Meson Clouds in Baryons

U. Zückert[1]

Institute for Theoretical Physics, Tübingen University
Auf der Morgenstelle 14, D-72076 Tübingen, Germany

## ABSTRACT

In the Nambu–Jona-Lasinio (NJL-) model baryons are considered either as quark-diquark-composites or as soliton configurations of mesons. The construction of a hybrid model, which possesses a dynamical interplay between both pictures, is presented. The Bethe-Salpeter equation for diquarks and the Faddeev equation for diquark-quark bound states in the background of a soliton configuration is solved.

At present there exist two generic classes of models to describe baryons. On the one hand there is the valence quark picture, which *e.g.* enters into nonrelativistic quark models, bag models or the description of baryons as quark-diquark bound states. In comparison, the soliton picture has successfully been applied to investigations of baryon properties. The soliton description results from large $N_C$-QCD studies and allows one to embed the useful concept of chiral symmetry and its spontaneous breaking straightforwardly. Therefore, low energy properties which are related to chiral symmetry, *e.g.* the matrix elements of the axial singlet current, are successfully described by soliton phenomenology. In contrast, quarks are seen experimentally in high energy scattering. From a technical point of view, the valence quark picture directly leads to the quantum numbers of a physical baryon whereas the soliton can only be interpreted as a baryon after collective quantization. These few examples indicate that despite their successes both approaches have only limited ranges of applicability. For that reason a unification of the two approaches is desirable. One such hybrid model is represented by the chiral bag model. It contains explicit quark degrees of freedom inside the bag while a chiral meson cloud dwells outside. According to the Chesire cat principle physical quantities should not depend on the radius of the bag, but the Chesire cat principle fails for axial matrix elements. Therefore, a hybrid model is needed which is completely determined by its dynamics. In this context the NJL-model is unique. This model makes possible a consistent unification by means of hadronization techniques [1] without any double counting of correlations, because it contains both, soliton solutions of meson fields [2] as well as diquark-quark bound states[3].

The realization of the rigorous self-consistent solution represents a complicated task. Nevertheless, an approximate evaluation of this hybrid baryon can be done within a four step procedure. As starting point we choose the NJL-model for two flavors with a pointlike interaction of color octet flavor singlet currents. The currents are rearranged by Fierz transforming into attractive meson ($\bar{q}q$ color singlet) and diquark ($qq$ color anti-triplet) channels. Using path integral hadronization techniques this quark theory has been converted into an effective theory of mesons, diquarks and baryons.

In the first step a static ground state solution in the absence of diquark and baryon fields is computed. Of course, this is nothing but the chiral soliton of the NJL model. For simplicity we restrict ourselves to the well-known hedgehog ansatz for the meson fields. The static energy of this soliton is given by $E_{sol} = E_{sea} + E_{val}$ where $E_{sea}$ corresponds to the part which results from the sea part of the action. The valence quark part is simply $E_{val} = 3\epsilon_{val}$ with $\epsilon_{val}$ the single particle energy of the valence quark state. The equation of motion is given by minimizing $E_{sol}$ with respect to the chiral angle $\Theta(r)$.

---

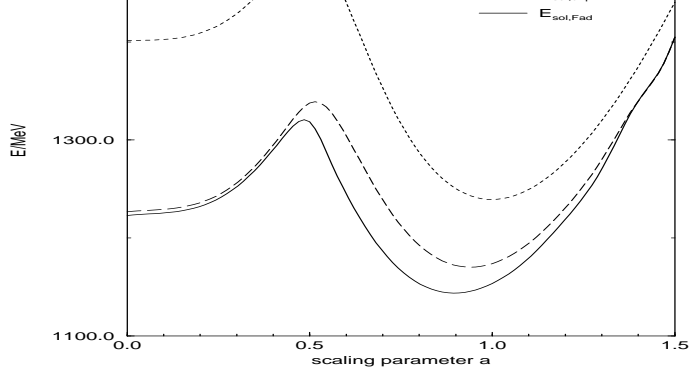

Figure 1: Comparison between the pure soliton energy $E_{sol}$ (dotted line), the soliton energy $E_{sol,diq}$ (dashed line) of the additive diquark-quark model and the energy $E_{sol,Fad}$ (solid line) of the soliton with diquark-quark correlations for constituent quark mass $m = 450$MeV.

Subsequently, we are interested in the behavior of diquarks in the background field of the soliton configuration. For this purpose the total action is expanded around the static soliton configuration up to quadratic order in the diquark fields yielding the Bethe–Salpeter equation for diquark fields in the soliton background (cf. [4]). We find that this equation possesses an energetically stable solution for an S-wave scalar diquark field with an effective diquark mass ($\omega_{diq} \approx 300$ MeV) strongly lowered through the presence of the soliton in comparison to the vacuum case ($\omega_{diq} \approx 760$ MeV). To have a self-consistent description of mesonic as well as diquark correlations we choose an one parameter family of soliton fields. For that reason, we scale the self-consistent solution in his softest mode - the radius - via $\Theta(r) = \Theta_{s.c.}(r/a)$. Now, the effect of this diquark mode on the soliton is considered in an additive model where two of the three valence quarks in the soliton are replaced by a bound diquark ($E_{sol,diq} = E_{sea} + \omega_{diq} + \epsilon_{val}$). The dependence of $E_{sol,diq}$ on the scaling parameter $a$ is shown in figure 1. We observe a new minimum for $a \approx 0.94$. This shows that the diquark correlations cause some shrinking of the soliton profile.

In the next step we integrate over the diquark fields. This induces in the action diquark propagators as well as interaction graphs for quarks and diquarks. In this infinite number of interaction graphs we neglect all self interactions which is nothing but the well-known ladder approximation. The Faddeev equation to calculate a bound state between quark and diquark interacting via quark exchange is given by minimizing the valence part of the action with respect to the baryon fields. If a bound state solution of the Faddeev equation $\Omega$ exists, a soliton with inclusion of diquark-quark interaction can be constructed by replacing the non-interacting quark and diquark with the quark-diquark bound state: $E_{sol,Fad} = E_{sea} + \Omega$. In figure 1 we recognize that the energy of this soliton configuration is again reduced. Also, the size of the soliton is shrinked further.

In conclusion, we have seen that a soliton with quark-quark as well as diquark-quark correlations is energetically favored without changing the shape of the soliton strongly. Because of the natural separation of the action into valence and sea parts the hybrid model of NJL is a suitable tool to separate valence and sea contributions of physical observables in baryons. But for such a calculation it is necessary to perform the last step in the construction of the baryon: the soliton has to be projected to physical quantum numbers which can be done with the help of the cranking method.

**Acknowledgments:** I would like to thank my colleagues R. Alkofer, H. Weigel who participated in the work underlying this talk. I am also grateful to H. Reinhardt for helpful discussions. Furthermore, I thank the Deutsche Forschungsgemeinschaft (DFG) for financial support for the Summer School.